\documentclass[prl,twocolumn,superscriptaddress,amsmath,showpacs]{revtex4} %superscriptaddress,preprintnumbers,amsmath,amssymb

\usepackage{graphicx}% Include figure files
\usepackage{dcolumn}% Align table columns on decimal point
\usepackage{bm}% bold math

\begin{document}

%\preprint{APS/123-QED}

\title{Photon delocalization transition in dimensional crossover in layered media}% Force line breaks with \\

\author{Sheng Zhang}
\affiliation{Department of Physics, Queens College, The City University of New York, Flushing, NY 11365, USA}
\author{Jongchul Park}%
\affiliation{Department of Physics, Queens College, The City University of New York, Flushing, NY 11365, USA}%
\author{Valery Milner}
\affiliation{Department of Physics, Queens College, The City University of New York, Flushing, NY 11365, USA}
\author{Azriel Z. Genack}%
\affiliation{Department of Physics, Queens College, The City University of New York, Flushing, NY 11365, USA}%

\date{\today}

\begin{abstract}
We report a crossover in optical propagation in nonuniform random layered media from localization towards diffusion as the interaction of the wave with the sample is transformed from one to three-dimensional. The crossover occurs at the point that the lateral spread of the wave equals the transverse coherence length in the transmitted speckle pattern. Delocalization is fostered as the sample thickness or lateral nonuniformity increases. 
\end{abstract}

\pacs{42.25.Dd, 42.25.Bs, 42.30.Ms }% PACS, the Physics and Astronomy
                             % Classification Scheme.

\maketitle
Layered media \cite{1} are ubiquitous in geological, biological, electronic, and photonic settings. Understanding transport in these largely one-dimensional structures embedded in three-dimensional space is challenging because of the critical role of dimensionality in wave propagation and localization. Both classical and quantum waves become exponentially peaked or localized \cite{2,3,4,5} in disordered samples when the number of times a wave winds its way through typical coherence volumes within the sample exceeds unity. The return to a point is reinforced by the constructive interference of waves following time reversed paths and is facilitated in low-dimensional systems which restrict the volume explored by the wave. As a consequence, localization can always be achieved in sufficiently large one and two-dimensional samples even when scattering is weak \cite{3}. In three dimensions, however, localization can only be realized when scattering is sufficiently strong that the mean free path, $\ell$, is substantially smaller than the wavelength, which may be expressed as, $k\ell <1$,\cite{6} where $k=2\pi/\lambda$  is the wavevector. 

Examples of one-dimensional localization abound. Localization has been observed for acoustic waves along a wire to which masses are randomly attached \cite{7}, microwave radiation in single-mode metallic waveguides with random dielectric inserts \cite{8}, and infrared radiation in single-mode fibers with random Bragg gratings \cite{9}. A one-dimensional description \cite{10} is also suitable in the case of plane wave illumination of an unbounded medium comprised of parallel layers. It has been used to describe the localization of electrons in semiconductor superlattices \cite{12} and photons in parallel dielectric layers of random thickness \cite{13,15,16}. Measurements of the scaling of average optical transmission, $\langle T(L)\rangle$, for a normally incident beam in an ensemble of random stacks of overhead transparencies \cite{15} and glass cover slips \cite{16} were in accord with 1D simulations. The sample thickness, $L$, is given in terms of the numbers of glass layers which alternate with air gaps. Transmission approached the asymptotic limit, $\langle T(L)\rangle \sim \exp(-L/2\xi)$, where $\xi$ is the calculated average exponential decay length of localized modes within the sample \cite{15,16}. Transmission in such samples is mediated by the excitation of states with single or multiple exponential peaks \cite{5} in the spatial intensity distribution \cite{8,17}. Localized modes in layered samples play a particularly important role in amplifying media since such modes are long-lived by virtue of their weak coupling to the boundaries. Low-threshold lasing was demonstrated in a stack of glass slides and dye sheets when the pump laser and emission spectrum overlapped localized modes near the center of the sample \cite{16}.

Though propagation and lasing in passive and active layered media has been extensively investigated, the impact of nonuniformity within the layers upon transport has not been reported. Instead, studies of waves in random layered media have focused on their localization perpendicular to presumed uniform layers. In this Letter, we report a crossover from localized towards diffusive propagation with increasing thickness and disorder in random layered media with nonparallel interfaces. Beyond the crossover point, transmission departs from 1D simulations and approaches an inverse rather than an exponential scaling. This reflects a continuous change in dimensionality of wave transport from one to three dimensions with increasing sample thickness. The crossover occurs because destructive interference, which results in localization in samples with uniform layers of random thickness, is washed out as wave trajectories spread beyond a coherence length in the transmitted speckle pattern due to transverse disorder.

We consider the nonuniformity in thickness within the layers. We studied transmission of a single frequency helium-neon laser at 633 $nm$ through stacks of 22-$mm^2$ glass slides with refractive index $n = 1.523$ and thicknesses in the range 125-135 $\mu m$. Samples are held in place by two rings of 18-mm inner diameter. Since the thickness of each glass slide and of the air gap between slides is not uniform, a normally incident beam is scattered off the normal direction to produce a speckled intensity pattern at the output. These speckle patterns can be imaged with a lens upon a CCD camera or scanned with an optical fiber probe leading to a photodiode detector (see \cite{EPAPS} for details). Examples of measured speckle patterns for samples with 1, 2, 20 and 80 slides are shown in Fig. 1. The degree of parallelism of the two faces of the slide can be ascertained from the fringe patterns of single slides [see Fig. 1(a)]. The fringes are generally nearly parallel to the sides of the slide. The fringe spacing, $a$, varies from 160 to 6800 $\mu m$ in a sample of 100 slides, indicating a variation of local wedge angle, $\theta$, from $1.5\times 10^5$ to $2.6\times 10^3$ rad, where $2na\theta \approx \lambda$ and $n=1.523$ is the refractive index of glass. The air gaps between slides are nonuniform because of deviations from flatness of the glass surfaces as well as because of occasional dust particles. This is reflected from speckle patterns generated by 2 slides [seen in Fig. 1(b)]. With increasing $L$, the speckle patterns at the output are randomized while the angular distribution of transmitted radiation broadens so that the scale of features in the speckle pattern shrinks.

\begin{figure}
\includegraphics[scale=0.9]{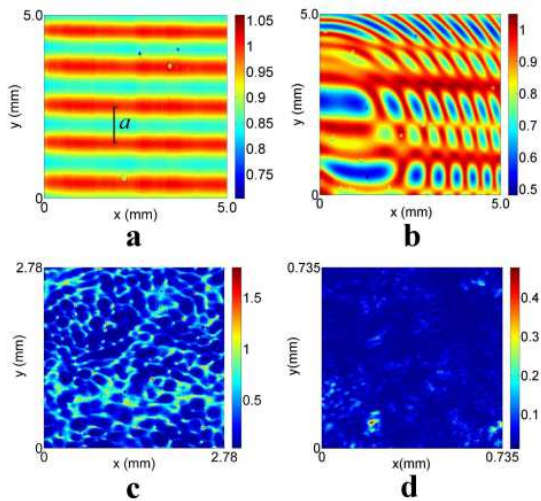}
\caption{\label{Fig1} (Color Online) (a) Example of a fringe pattern generated by a single slide. (b) Typical speckle pattern for two slides and intervening air gap. (c) and (d) Speckle patterns generated by samples with 20 and 80 slides, respectively. The local transmission coefficient indicated by the colorbar may exceed unity due to the interference of waves with wavevector components in the layer plane.}
\end{figure}

\begin{figure}
\includegraphics[scale=0.9]{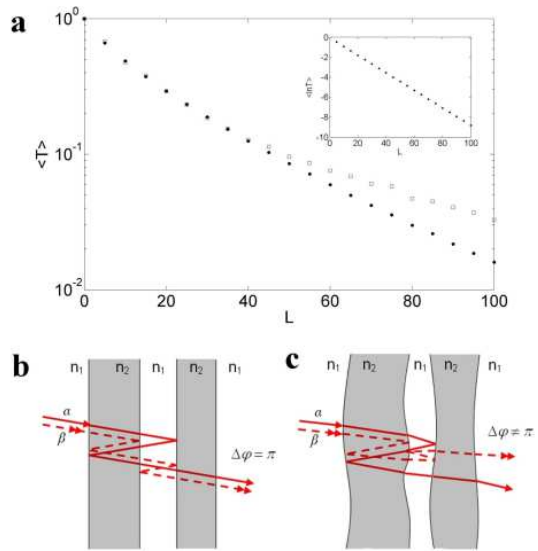} 
\caption{\label{Fig2} (Color Online) (a) Semi-logarithmic plot of measurements (red squares) and 1D simulations (black dots) of $\langle T\rangle$ versus number of glass slides, L. Simulations of $\langle \ln T\rangle$ are shown in the inset. (b) and (c) Schematic of interference between partial waves following two trajectories, $\alpha$ and $\beta$, which pass through the same layers an equal number of times, in samples with parallel and nonparallel layers, respectively. }
\end{figure}

The average transmission of a 500-$\mu m$-wide collimated beam directed normal to the glass slides was measured using an integrating sphere. Transmission was averaged by translating the sample over a 100-$mm^2$ area for 10 different stacks of slides. Measurements are plotted with red squares in Fig. 2(a) and compared with 1D simulations for an ensemble of configurations shown as black dots. In the simulations, transmission and reflection for each layer is represented by a   transfer matrix and transmission for the entire structure can be obtained from the product of these matrices. Simulations of $\langle T(L)\rangle$ fall exponentially in the limit of large $L$. The decay length is predicted to equal $2\xi$, which is equal to the thickness of a stack of 22 slides. $\xi$ is also the exponential decay length $\langle \ln T(L)\rangle$ [see inset in Fig. 2(a)].  Measurements of $\langle T(L)\rangle$ are in agreement with simulations up to $L = 40$ but fall more slowly than simulations for larger $L$.

The departure of measurements of $\langle T(L)\rangle$ from 1D simulations can be understood by comparing the superposition of corresponding rays in samples with parallel and nonparallel interfaces shown schematically in Figs. 2(b) and 2(c). The two wave trajectories $\alpha$ and $\beta$ in the sample with parallel interfaces shown in Fig. 2(b) pass through each slide the same number of times and are therefore of equal length. However, since light reflects from a higher index medium in one of the two additional reflections in path $\beta$, the partial waves for the two paths are out of phase by $\pi$ rad and interfere destructively. As the number of layers increases, the relative weight of such pairs of out of phase trajectories increases leading to an exponential decrease of $\langle T(L)\rangle$. \cite{15}

In a sample in which the interfaces are not parallel, the trajectories corresponding to those in Fig. 2(b) are distorted as shown schematically in Fig. 2(c). An additional phase difference between the two partial waves accumulates since trajectories cross the layer at different points at which the layer thicknesses differ and the angle between trajectories are no longer equal. The phase difference between such pairs of trajectories is thereby increasingly randomized as the spatial and angular spread of the beam increases with increasing number of slides or local wedge angles. The reduced cancellation of transmission of such paired trajectories leads to a slower decay of $\langle T(L)\rangle$. In the limit in which the correlation between such pairs of partial waves vanishes, the falloff of $\langle T(L)\rangle$ becomes diffusive and transmission falls as $1/L$ \cite{20}.

The above considerations make it plain that nonuniformity within the layers reduces the impact of localization on transmission. The suppression of longitudinal localization depends upon the relationship between typical displacements within the plane between trajectories starting at the same point and the field coherence length, each of which is influenced by disorder within the layers. The coherence length is directly exhibited in the intensity speckle pattern and is inversely proportional to the width of angular spread of the transmitted beam \cite{21}. Wave localization is essentially one dimensional only as long as the area explored by a wave incident at a point is smaller than the coherence area of the field. Thus, one-dimensional localization breaks down once the sample is thick enough that the characteristic length of the transverse spread of the wave, $\sigma_\perp$ , equals the field correlation length in the plane, $d_\perp$, $\sigma_\perp=d_\perp$. These lengths along a single direction are shown schematically in Fig. 3(a). 

\begin{figure}
\includegraphics[scale=0.9]{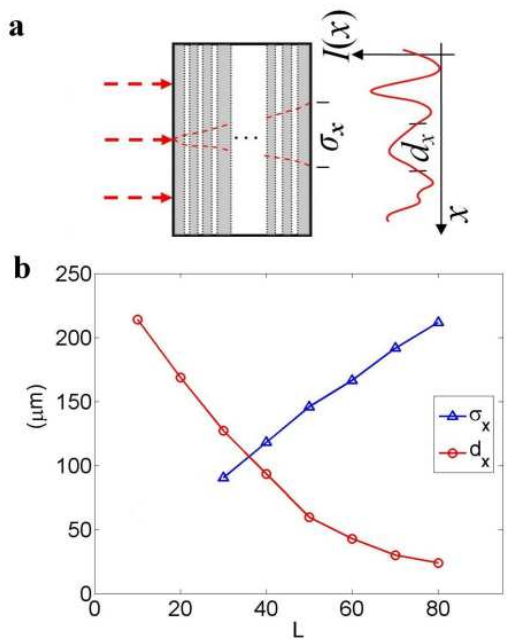} 
\caption{\label{Fig3} (Color online) (a) Schematic of the average transverse spread, $\sigma$, and the speckle size, $d$, along $x$-direction at the output plane for an incident plane wave. (b) Measurement of $\sigma_x$ and $d_x$ versus $L$. Their crossing at $L\approx 35$ is consistent with the departures of $\langle T(L)\rangle$ from results of 1D simulations beginning at $L=35$ [seen in Fig. 2(a)]. }
\end{figure}

The field correlation length along the $x$-direction, $d_x$, for example, can be determined from the correlation function of the field on the output surface,  $\Gamma(\Delta x)$. This in turn is the Fourier transform of the ensemble averaged angular distribution \cite{21} measured in the far-field, known as the specific intensity, $\langle I(\theta_x)\rangle$ \cite{EPAPS}. The correlation lengths are taken to be twice the length in which Re$\{\Gamma(\Delta x)\}$ decays to half its maximum value.

The ensemble average of the spread of an incident beam with intensity profile, $I_{in}(x^\prime,y^\prime)$, to produce the intensity distribution at the output surface, $\langle I_{out}(x,y)\rangle$ may be expressed in terms of the spread function, $P_{in}(x-x^\prime, y - y^\prime)$, giving $\langle I_{out}(x,y)\rangle = \int\int I_{in}(x^\prime,y^\prime) P_{in}(x-x^\prime, y - y^\prime) dx^\prime dy^\prime$. $P_{in}(x-x^\prime, y - y^\prime)$ is similar to the point spread function in three-dimensional diffusive systems but differs in that it depends upon the wavevector distribution at the input. Therefore, standard methods employed to determine the point spread in three-dimensional random samples such as the direct measurement of the beam profile due to a strongly focused incident beam \cite{22} or the measurement of the intensity correlation function in the far field as a function of sample angle \cite{23} cannot be applied in layered samples. Direct measurement of $I_{in}(x^\prime,y^\prime)$ and $\langle I_{out}(x,y)\rangle$ are made by imaging the wave onto a CCD camera (see \cite{EPAPS} for details). Integrating over $y$, for example, gives, $I_{in}(x) = \int I_{in}(x,y)dy$. Defining $\sigma^2_{x_{in}}$, $\sigma^2_{x_{out}}$ and $\sigma^2_{x}$ as the variances based on the functions, $I_{in}(x)$, $\langle I_{out}(x)\rangle$ and $P(\Delta x)$, respectively, and taking the origin as the center of the incident beam, gives, $\sigma^2_{x_{in}}=\int_{-\infty}^\infty I_{in}(x)x^2dx$, and $\sigma^2_{x} = \sigma^2_{x_{out}} - \sigma^2_{x_{in}}$. The variance   so obtained can be used to characterize the spread of the wave in the $x$-direction.

The variations of the widths of the intensity spread functions and the speckle size along the $x$-direction as a function of the number of slides are shown in Fig. 3(b). Results for $\sigma$ and $d$ along the $y$-direction are very close to those along the $x$-direction. The measurements of $\sigma$ and $d$ allow us to determine the effective number of transverse modes involved in transmission of the wave over the area over which the wave spreads, $\sigma_x \sigma_y$, $N = N_xN_y$, where, 
%\begin{equation}
\[ N_x = \left\{ 
     \begin{array}{cc}
       1,& \sigma_x/d_x <1; \\
       \sigma_x/d_x, & \sigma_x/d_x >1.
     \end{array}
     \right. \]
%\end{equation}
and $N_y$ is similarly defined. When $N = 1$, wave propagation is essentially one-dimensional. Only a single polarization component of the wave is considered since transmission is highly polarized even in the thickest samples. The crossing of the curves for $\sigma$ and $d$ marks a crossover from one to three-dimensional transport and a transition from localization to diffusion. Such a crossing occurs at $L\approx 35$ for both the $x$ and $y$ directions [Fig. 3(b)]. Beyond this thickness, $\langle T(L)\rangle$ departs from 1D simulations [Fig. 2(a)].

\begin{figure}
\includegraphics[scale=0.9]{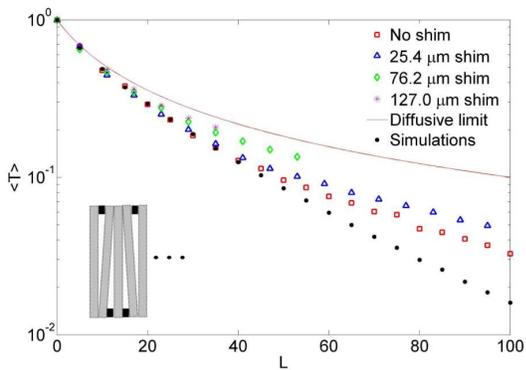}% Here is how to import EPS art
\caption{\label{Fig4} (Color Online) Comparison of $\langle T(L)\rangle$ for samples with different additional wedge angles. Thin shims with different thicknesses are inserted between layers to introduce average wedge angles of $0.073^\circ$, $0.218^\circ$ and $0.364^\circ$, respectively, into the air gaps.  The side of the slide in which the shims are placed is alternated so that the average angle of the slides is not changed. The total transmission could only be measured up to thicknesses at which the beam spread does not approach the edges of the slides.}
\end{figure}

Since transverse disorder leads to both an increased spread of the wave and to a drop in the coherence length, we expect that $\langle T(L)\rangle$ will depart from 1D simulations when the degree of nonparallism of the layers increases. This is confirmed in measurements in samples created by inserting narrow metal shims at alternating edges of the glass slides as shown in Fig. 4. The curve in Fig. 4 is the calculation for photon diffusion utilizing the intensity reflection coefficient at the air/glass interface, $R = [(n-1)/(n+1)]^2 = 0.043$ . The falloff of $\langle T(L)\rangle$ approaches the diffusive limit as the wedge angle increases.

The scaling of transmission in layered samples differs from scaling observed in samples in which wave propagation is of fixed dimensionality \cite{20,25,26}. In such samples, the wave, once localized, remains localized, and the scale dependent conductivity or diffusion coefficient \cite{4,25,26,27} decrease continuously with sample thickness. This is in contrast to wave delocalization observed here in layered media. Unlike propagation in isotropic two or three-dimensional random media, for which the angular distribution of transmission is independent of thickness for $L>\ell$, the angular distribution of the wave in layered media is highly directional and broadens with sample thickness and with depth into the sample. 

Highly anisotropic angular distributions are also found in samples in which the index of refraction is uniform in the longitudinal direction but disordered in the transverse directions \cite{28,29,30} The small values of $k_\perp$ in that case leads to localization in the transverse plane in relatively short distances even when, $k\ell_\perp\gg 1$, where $\ell_\perp$ is the transverse mean free path \cite{29}. Such transverse localization stands in contrast to longitudinal delocalization in random layered samples which is exhibited once the spread of the wave exceeds the transverse coherence length. 

In conclusion, the crossover from localized to diffusive transport in layered media demonstrates the critical role of dimensionality in transport in a class of samples which occurs widely in nature and in photonics and electronic microstructures.

We thank Victor Kopp for the 1D simulation program, Samuel Gillman for experimental assistance, and Howard Rose for the sample holder. This research was sponsored by the National Science Foundation under grant no. DMR-0538350. V.M. is presently at the Department of Chemistry and the Laboratory for Advanced Spectroscopy and Imaging Research, University of British Columbia, Vancouver, BC, V6T 1Z1, Canada.

\end{document}